\documentclass[aps,prl,preprint,groupedaddress,showpacs]{revtex4}

\usepackage{bm} 
\usepackage{amsmath}
\usepackage{amsfonts}
\usepackage{amssymb} 
\usepackage{longtable}
\usepackage{epsf} 
\usepackage[dvips]{epsfig,color} 
\usepackage{version}
\usepackage{graphicx}
\usepackage{psfrag}

\newcommand{\opencircle}{\mbox{\Large$\circ$}}
\newcommand{\fullcircle}{\mbox{{\Large$\bullet$}}}
\newcommand{\opensquare}{\mbox{$\Box$}}
\newcommand{\fullsquare}{\mbox{$\blacksquare$}}

\newcommand{\dashedline}{\protect\rule[2pt]{3pt}{1pt} \!\protect\rule[2pt]{3pt}{1pt} \!\protect\rule[2pt]{3pt}{1pt}}
\newcommand{\fullline}{\protect\rule[2pt]{15pt}{1pt}}

\newcommand{\rmd}{\mathrm{d}}
\newcommand{\red}[1]{{\color{red}#1}}
\newcommand{\green}[1]{{\color{green}#1}}
\newcommand{\blue}[1]{{\color{blue}#1}}

\begin{document}
\title{Entropy-driven enhanced self-diffusion in confined reentrant
  supernematics} 
\author{Marco G. Mazza$^{\dagger}$, Manuel
  Greschek$^{\dagger}$, Rustem Valiullin$^{\ast}$, J\"org K\"arger$^{\ast}$, and Martin Schoen$^{\dagger,\ddagger}$} 

\affiliation{
  $^{\dagger}$Stranski-Laboratorium f\"ur Physikalische und
  Theoretische Chemie, Technische Universit\"at Berlin, Stra{\ss}e des
  17. Juni 135, 10623 Berlin, Germany\\
  $^{\ast}$Institut f\"ur Experimentelle Physik I, Universit\"at
  Leipzig, Linn{\'e}str. 5, 04103 Leipzig, Germany\\
  $^{\ddagger}$Department of Chemical and Biomolecular Engineering, North Carolina State University, 911 Partners Way, Raleigh, NC 27695,
  U.S.A.
  }

\date{\today}
\begin{abstract}
  We present a molecular dynamics study of reentrant nematic phases using the
  Gay-Berne-Kihara model of a liquid crystal in nanoconfinement. At densities
  above those characteristic of smectic A phases, reentrant nematic phases
  form that are characterized by a large value of the nematic order parameter
  $S\simeq1$. Along the nematic director these ``supernematic'' phases exhibit a remarkably high
  self-diffusivity which exceeds that for ordinary, lower-density nematic
  phases by an order of magnitude. Enhancement of self-diffusivity
  is attributed to a decrease of rotational
  configurational entropy in confinement. Recent developments in the pulsed field gradient NMR technique are shown to provide favorable conditions for an experimental confirmation of our simulations.
\end{abstract}
\maketitle
In the context of phase transitions ``reentrancy'' refers to a nonmonotonic variation of an order parameter with the thermodynamic field driving the transition. Reentrancy is ubiquitous in the physics
of thermal many-particle systems. It arises under quite disparate
physical conditions encountered, for example, in quantum gases \cite{kleinert04}, two-dimensional
charged colloids \cite{bechinger00}, or
relativistic scalar field models \cite{pinto05}. As far as soft matter is concerned reentrancy has been
reported for self-assembled supramolecular structures
\cite{osaka07,dudowicz09}, wetting phenomena at oleophilic surfaces
\cite{ramos10}, and novel discotic and calamitic liquid crystals
\cite{szydlowska08}. In fact, since the first observation of reentrant nematic
(RN) phases in a seminal paper by Cladis \cite{cladis75} reentrancy in liquid crystals seems to have received most of the attention.
This is most likely because of the abundance of phases exhibited by these
materials. For example, reentrant phase transitions have been reported for the
isotropic phase of mixtures of discotic liquid crystals \cite{lee89}, the
ferroelectric transition in syn- and anticlinic smectic C phases
\cite{pociecha01}, the cholesteric-to-blue phase transition in chiral liquid
crystals \cite{heppke90}, and for nematic (N) phases \cite{sigaud81}. 

Despite
the variety of systems and thermodynamic conditions under which reentrancy in
liquid-crystalline materials arises comparatively little attention has been
paid to the dynamics of reentrant phase transitions. For example, distinct differences in the molecular dynamics in the N and RN phases can be concluded from corresponding changes in the nuclear magnetic relaxation times reported in Refs.~\citealp{dong8182, miyajima84,bharatam99}.

Whereas most earlier work on reentrancy of phase transitions in liquid
crystals is experimental in nature comparatively little attention has been
paid to this fascinating phenomenon from a theorist's point of view. The most
recent theoretical study employs isothermal-isobaric and canonical ensemble
Monte Carlo (MC) simulations to investigate the nature of the smectic
A(smA)-RN phase transition for a bulk system of hard ellipsoids with
square-well attraction \cite{demiguel05} where earlier theoretical studies are
briefly reviewed, too. Unfortunately, the model employed in
Ref.~\citealp{demiguel05} is somewhat artificial in assuming that the
ellipsodal molecules are always oriented in a perfectly parallel fashion such
that all rotational degrees of freedom are always ``frozen'' irrespective of
the thermodynamic conditions. Therefore, this study seems only of limited use
to elucidate properties of RN phases at a molecular level. Moreover, the
authors do not consider dynamic features of RN phases.

Therefore, we show here
that the smA-RN phase transition causes a dramatic increase in the
self-diffusion of the molecules in the direction of the nematic director
$\widehat{\bm{n}}$. Our model system consists of soft spherocylinders
interacting via the so-called Gay-Berne-Kihara pair potential \cite{martinez05}; specifically,
we use the GBK(6,5,2,1) version of the model in the notation of
Ref.~\citealp{martinez05}. Data will be presented for a system of spherocylinders
confined to a slit-pore with structureless walls, separated by a distance
$s_z$ along the $z$-axis. The fluid-solid interaction is described by the
surface-averaged potential \cite{schoen07}
\begin{equation}\label{eq:ufs}
u_{\mathrm{fs}}=4\epsilon_{\mathrm{fs}}\rho_{\mathrm{s}}\sigma^2
\left[
\left(\frac{\sigma}{d_{\mathrm{w}}}\right)^{10}-
\left(\frac{\sigma}{d_{\mathrm{w}}}\right)^4
g(\widehat{\bm{u}})
\right].
\end{equation}
In Eq.~(\ref{eq:ufs}), $\epsilon_{\mathrm{fs}}$ determines the depth of the
attractive well, $\rho_{\mathrm{s}}\sigma^3=2^{-1/3}$ is the areal density of
a layer of substrate atoms, $d_{\mathrm{w}}$ is the minimum distance of a
spherocylinder from either substrate, and $g\left(\widehat{\bm{u}}\right)$ is
the so-called ``anchoring function'' where the unit vector $\widehat{\bm{u}}$
specifies the orientation of a molecule. The anchoring function is introduced
to discriminate energetically specific orientations of molecules with respect
to the substrate plane following a suggestion of Steuer {\em et al.}
\cite{steuer04}. Here,
$g\left(\widehat{\bm{u}}\right)=u_{\mathrm{x}}^2+u_{\mathrm{y}}^2$ such that
an orientation parallel with the substrate plane is favored \cite{greschek10};
throughout this work $s_{\mathrm{z}}=19\sigma$ is used.

We employ both extensive isothermal-isobaric MC and microcanonical molecular
dynamics (MD) simulations to locate the smA-RN phase transition and to study
changes in mass transport accompanying that transition. The main purpose of
employing MC is to provide suitably equilibrated starting configurations for
subsequent MD simulations. In addition, MC is used to independently verify the
correctness of the MD simulations through a comparison of equilibrium
properties obtained in both types of simulations. In MC we use a standard algorithm \cite{schoen99} but allow the
side lengths of the computational cell to vary independently to preserve the
in-plane isotropy of the pressure tensor even in highly ordered confined
phases; MD simulations are based on an implementation of the velocity Verlet
algorithm \cite{ilnytskyi02}. We employ the customary dimensionless units of
$\sigma$ for length, $\epsilon/k_{\mathrm{B}}$ for temperature $T$, and
$(\sigma^2m/\epsilon)^{1/2}$ for time $t$, where $m=1$ is the spherocylinder
mass. Values for $\sigma$ and $\varepsilon$ are taken from
Ref.~\citealp{martinez05}. Our simulations comprise $N=1000$ molecules of
length $L=6$. Interactions are truncated beyond a minimum distance
$d_{\mathrm{m}}=3$. In MD the starting configuration is further equilibrated
using a velocity rescaling scheme to the desired $T$ at constant volume.
Throughout this work we employ a time step of $\Delta t=10^{-4}$ to integrate
the equations of motion.

\begin{figure}
\begin{center}
\psfrag{yz}[][l]{$S$, $\Lambda$}
\psfrag{xz}[c]{$\rho$}
\includegraphics[scale=0.25]{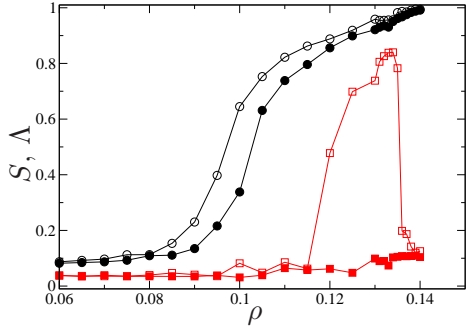}
\end{center}
\caption{\small (Color online) Plots of $S$ (\opencircle,\fullcircle) and
  $\Lambda$ (\red{\opensquare},\red{\fullsquare}) as functions of number
  density $\rho$ for confined liquid crystals. Open and filled symbols refer
  to $T=4.0$ and $6.0$, respectively.}\label{fig1}
\end{figure}

\begin{figure}
\begin{center}
\includegraphics[scale=0.16]{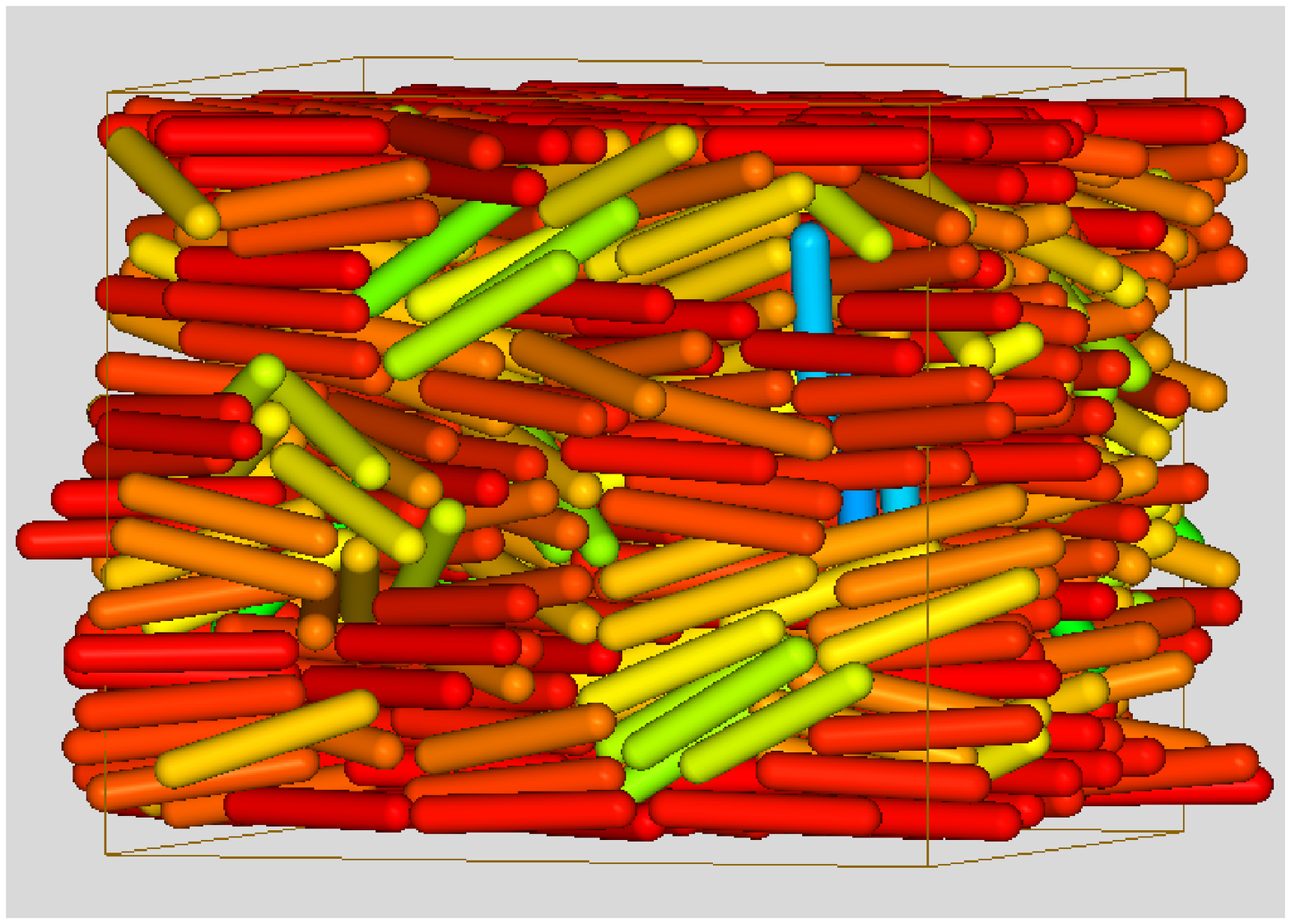}
\includegraphics[scale=0.16]{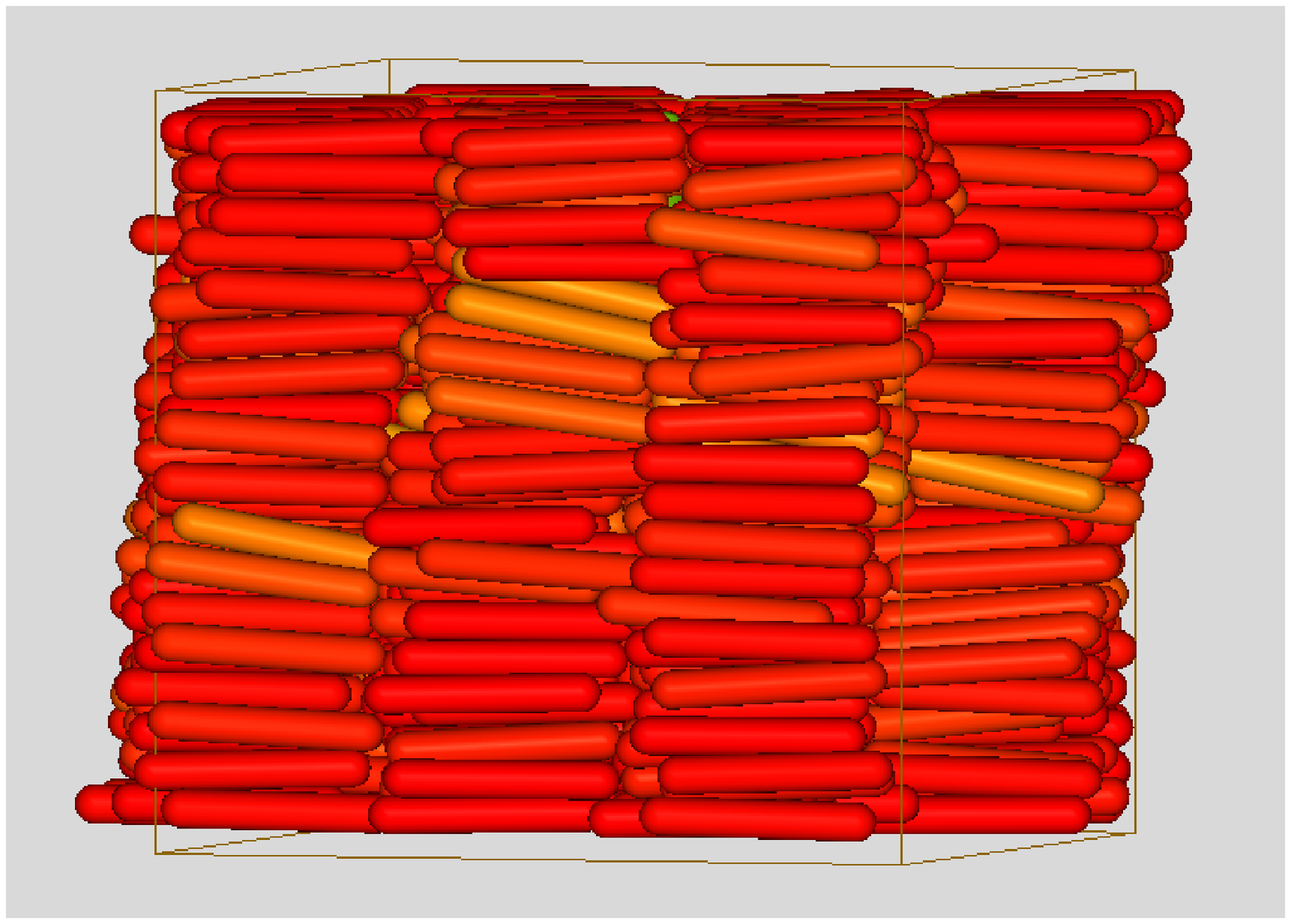}
\includegraphics[scale=0.16]{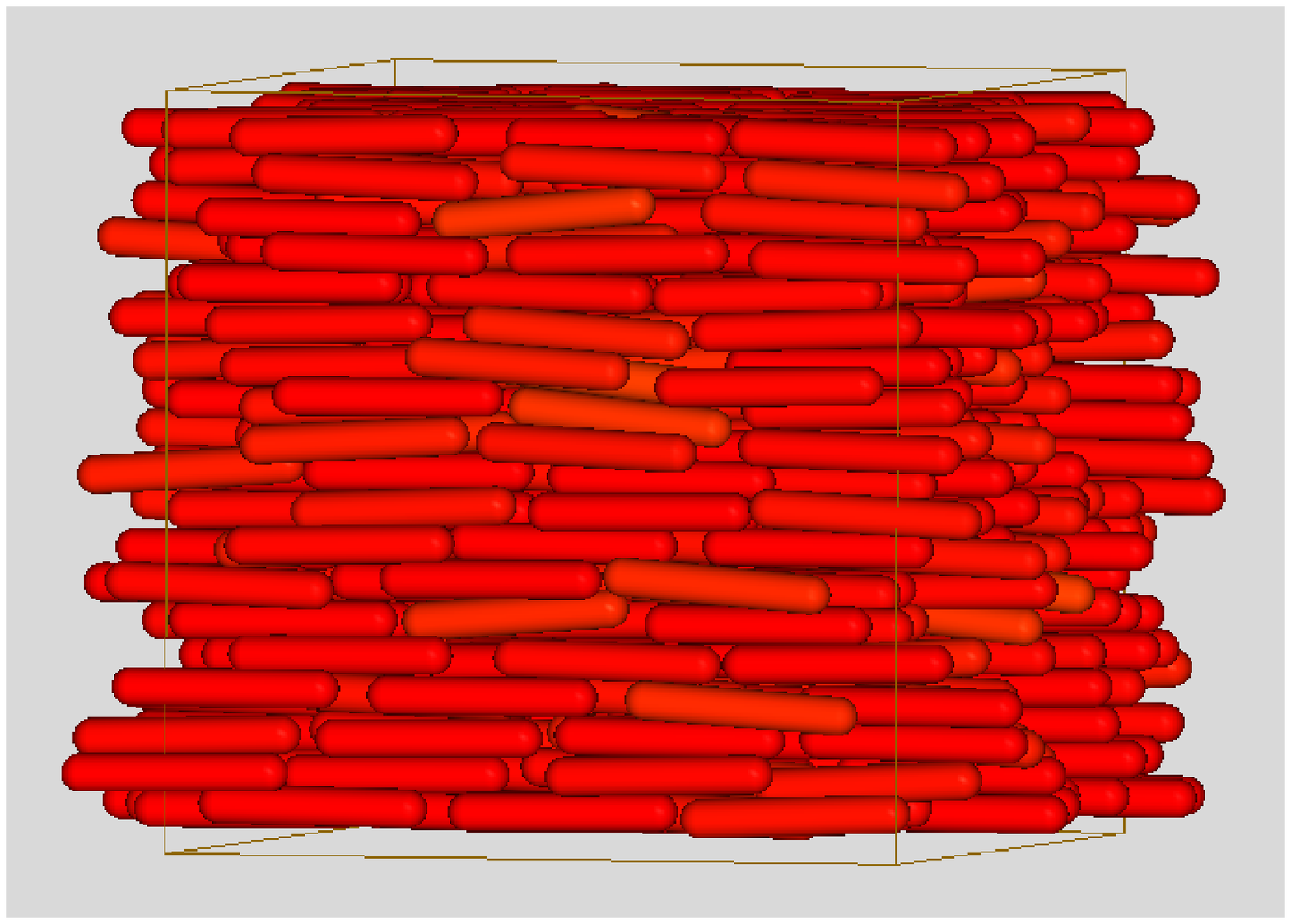}
\end{center}
\caption{\small (Color online) Configuration ``snapshots'' from MD; 
  $\rho=0.110$ (left, N), $0.133$ (middle, smA), $0.139$ (right, RN), $T=4.0$. Color code: $\widehat{\bm{u}}_i\cdot\widehat{\bm{n}}=0$ \blue{\fullline}, $\widehat{\bm{u}}_i\cdot\widehat{\bm{n}}=1$ \red{\fullline}, $i=1,\ldots,N$.}\label{fig2}
\end{figure}

To characterize the degree of nematic order we follow
established protocols and compute the nematic order parameter
$S$ as the ensemble average of the largest eigenvalue of the instantaneous
alignment tensor \cite{maier59}
\begin{equation}
\mathbf{Q}\equiv
\frac{3}{2N}
\left(
\sum\limits_{i=1}^N
\widehat{\bm{u}}_i\otimes\widehat{\bm{u}}_i
-\frac{1}{3}\mathbf{1}
\right)
\end{equation}
where ``$\otimes$'' denotes the direct product and $\mathbf{1}$ is the unit
tensor. Hence, $\mathbf{Q}$ is real, symmetric, and traceless. Its
eigenvector associated with the largest eigenvalue corresponds to
$\widehat{\bm{n}}$. The layering characteristic of smA phases is
quantitatively described through the function \cite{steuer04}
\begin{equation}
\lambda\left(d\right)\equiv
\left\langle
\left|
\frac{1}{N}\sum_{j=1}^N
\exp\left[
\frac{2\pi i\left(\mathbf{r}_j\cdot\widehat{\bm n}\right)}{d}
\right]
\right|
\right\rangle
\end{equation}
where $d$ is the spacing between layers and the smectic order parameter
$\Lambda$ is defined as the maximum of $\lambda\left(d\right)$ in the interval
$\left[L-\zeta,L+\zeta\right]$ where $\zeta=0.05$. Focusing on $T=4.0$ first,
plots in Fig.~\ref{fig1} show that $S$ is rather small up to a number density
$\rho\lesssim0.09$ indicating that the confined fluid is in its isotropic
phase. Beyond $\rho\gtrsim0.09$, $S$ rises steeply assuming values
characteristic of nematic phases ($S\gtrsim0.4$) and levels off as $\rho$
increases. At the highest densities considered $S\simeq1$ which reflects a
nearly perfect orientation of molecules. Over the same range of densities,
$\Lambda$ is very small up to values of $\rho\simeq0.115$, indicating that the
confined fluid does not form any smectic layers. As the density increases
further smectic layers are forming revealed by $\Lambda\gtrsim0.5$.
Interestingly, as the density keeps increasing beyond $\rho\simeq0.135$,
$\Lambda$ drops to a low value of about $0.1$ whereas $S$ increases further towards
its maximum value of about $1$. Hence, $\rho\simeq0.135$ demarcates the
formation of a RN phase.  The structural characteristics of confined phases
pertaining to the stable regime of N, smA, and RN phases is illustrated by
``snapshots'' of configurations generated in MD (see Fig.~\ref{fig2}). Notice
the presence of distinct molecular layers in the smA state which disappear
once the RN state forms. Comparing snapshots for typical N and RN phases it is
apparent from Fig.~\ref{fig2} that the latter exhibit much more orientational
but roughly the same positional order compared with the former.

\begin{figure}
\begin{center}
\psfrag{yz}[][l]{$\langle\Delta r_{\parallel}^2(\tau)\rangle_t$}
\psfrag{xz}[c][c]{$\tau$}
\includegraphics[scale=0.25]{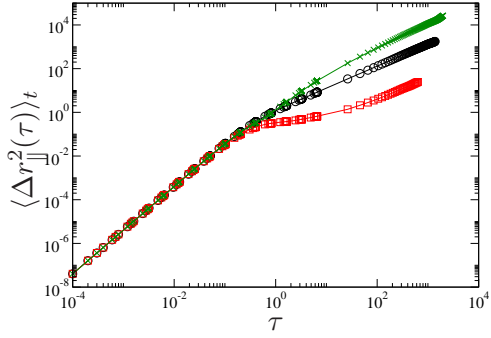}
\end{center}
\caption{\small (Color online) MSD's in the confined N (\opencircle,
  $\rho=0.110$), smA (\red{\opensquare}, $\rho=0.133$), and RN phase
  (\green{$\times$}, $\rho=0.139$) at $T=4.0$.}\label{fig3}
\end{figure}

\begin{figure}
\begin{center}
\psfrag{yz}[][l]{$D_{\parallel}$}
\psfrag{xz}[c][c]{$\rho$}
\includegraphics[scale=0.25]{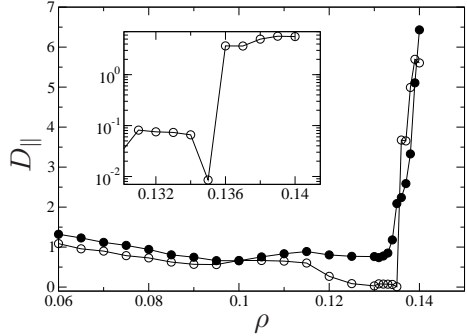}
\end{center}
\caption{\small $D_{\parallel}$ [see Eq.~(\ref{eq:dpara})] as a function of
  $\rho$ for the confined liquid crystal; (\opencircle) $T=4.0$, (\fullcircle)
  $T=6.0$. Inset is a magnification of the $T=4.0$ isotherm around the
  smA--RN transition.}\label{fig4}
\end{figure}

The key result of our study concerns the self-diffusion of molecules in the direction of their orientation. To that end we introduce the parallel mean-square displacement (MSD)
\begin{equation}\label{eq:msd}
\left\langle
\Delta r_{\parallel}^2\left(\tau\right)
\right\rangle_t\equiv
\frac{1}{N}
\left\langle
\sum\limits_{i=1}^N
\left\{
\widehat{\bm{u}}_i\cdot
\left[
\bm{r}_i\left(t+\tau\right)-
\bm{r}_i\left(t\right)
\right ]
\right\}^2
\right\rangle_{t}
\end{equation}
The angular brackets indicate an average over the $N$ molecules and a
sufficiently large number of time origins referred to by subscript ``$t$''. In
Fig.~\ref{fig3} we present plots of the MSD for typical N, smA, and RN phases
where different time regimes can be identified. In the double-logarithmic
representation short-time slopes of the MSD's exceed their long-time
counterparts on account of the initial ballistic motion of molecules in all
three phases. As expected, the MSD's are indistinguishable for $\tau\lesssim0.1$.
In the limit of large $\tau$ the time dependence of the MSD's corresponds to
diffusive motion, that is the MSD's depend linearly on $\tau$. The MSD for the
smA phase exhibits a plateau at intermediate times which eventually gives way
to diffusive motion at long times. The plateau reflects the presence of
smectic layers that hinder molecular motion in the direction perpendicular to
the plane of adjacent smectic layers (see Fig.~\ref{fig2}). In the limit $\tau\to\infty$ reliable values of
$D_{\parallel}$ can be extracted from the plots in
Fig.~\ref{fig3} via the Einstein relation \cite{allen90,loewen99}
\begin{equation}\label{eq:dpara}
D_{\parallel}=
\lim\limits_{\tau\to\infty}
\frac{1}{2\tau}
\left\langle
\Delta r_{\parallel}^2\left(\tau\right)
\right\rangle_t
\end{equation}
A comparison of plots in Figs.~\ref{fig1} and \ref{fig4} reveals that in the N
phase, $D_{\parallel}$ is small but nonzero. It attains a nearly vanishing
value in the smectic phase on account of layer formation (see Fig.~\ref{fig2})
which blocks mass transport efficiently in the direction of the layer normal
\cite{allen90}. Most importantly, however, compared with the smA phases
$D_{\parallel}$ increases dramatically by several orders of magnitude when the
RN phases become thermodynamically stable (i.e., for $\rho\gtrsim0.135$, see
inset in Fig.~\ref{fig4}). Because of the unusually large $D_{\parallel}$ in
conjunction with high nematic order ($S\simeq1$, see Fig.~\ref{fig1}) we call
these high-density nematic phases as ``supernematic''. One also notices by
comparing data for different $T$ in Fig.~\ref{fig4} that both data sets show a
similar large increase of $D_{\parallel}$ beyond a certain density threshold.
However, the plot for $T=6.0$ exhibits nonzero values of $D_{\parallel}$ over
a density range where the corresponding curve for the lower $T=4.0$ drops to
zero. This is due to the absence of smA phases at $T=6.0$ (see
Fig.~\ref{fig1}). Interestingly, the experimentally determined phase diagram
presented by Guillon {\em et al.} also suggests absence of intermittent smA
phases and a continuous transition from N to RN phases under suitable
thermodynamic conditions \cite{guillon78}. Nevertheless, the increase in
orientational order for $\rho\gtrsim0.135$ and $T=6.0$ causes $D_{\parallel}$
to increase equally strongly. Hence, at sufficiently high $T$ one may observe
supernematic features {\em without} reentrancy.

\begin{figure}
\begin{center}
\psfrag{yz}[][l]{$D_{\parallel}$}
\psfrag{xz}[c]{$\sigma_{\mathrm{rc}}^{-1}$}
\includegraphics[scale=0.25]{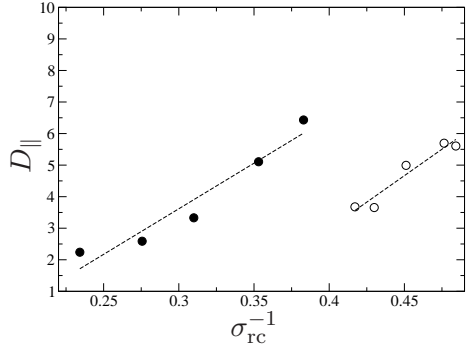}
\end{center}
\caption{\small $D_{\parallel}$ as a function of $\sigma_{\mathrm{rc}}^{-1}$ for $T=4.0$ (\opencircle) and $T=6.0$ (\fullcircle). The linear fit (\dashedline) is rationalized in the text.}\label{fig5}
\end{figure}

The apparent dramatic increase of self-diffusivity in the RN phase can be
attributed to a loss of rotational configurational entropy
$\mathcal{S}_{\mathrm{rc}}$ due to both increase in density and presence of
solid surfaces. We rationalize this by assuming a characteristic time interval
$\tau_{\parallel}$ associated with the onset of diffusive motion in the
direction of $\widehat{\bm{n}}$ such that
$D_{\parallel}\propto1/\tau_{\parallel}$. For $\tau\ge\tau_{\parallel}$ the
probability $\mathcal{P}$ that a molecule has traveled a distance $\Delta
r_{\parallel}$ from its origin at $\tau=0$ in the direction of $\widehat{\bm{n}}$
should then also be inversely proportional to $\tau_{\parallel}$. Intuitively,
one expects $\mathcal{P}\left(\Delta r_{\parallel}\right)$ to be larger if the
alignment of molecules with $\widehat{\bm{n}}$ is more pronounced on average,
that is the larger $S$ is. However, a larger value of $S$ implies a lower
rotational configurational entropy $\mathcal{S}_{\mathrm{rc}}$ such that
$\mathcal{P}\propto\exp\left(-\mathcal{S}_{\mathrm{rc}}/k_{\mathrm{B}}\right)$
using standard statistical-physical reasoning \cite{landau80}. This then
suggests
$D_{\parallel}\propto\exp\left(-\mathcal{S}_{\mathrm{rc}}/k_{\mathrm{B}}\right)$.
We estimate $\mathcal{S}_{\mathrm{rc}}$ via
\begin{equation}
\mathcal{S}_{\mathrm{rc}}=-k_{\mathrm{B}}
\int\rmd\theta \,P(\theta)\ln P(\theta)
\end{equation}
where $P(\theta)$ is the distribution of angles
$\cos\theta_i=\widehat{\bm{u}}_i\cdot\widehat{\bm{n}}$. Assuming $P(\theta)$
to be Gaussian with a standard deviation of $\sigma_{\mathrm{rc}}$ it is easy
to verify that $\mathcal{S}_{\mathrm{rc}}\propto\ln\sigma_{\mathrm{rc}}$. A
similar relation was obtained for the configurational entropy of a
macromolecule in Ref.~\citealp{karplus81} assuming a Gaussian distribution of
relevant internal degrees of freedom. Hence, the above line of arguments
suggests $D_{\parallel}\propto\sigma_{\mathrm{rc}}^{-1}$ which is supported by plots in
Fig.~\ref{fig5}.

In summary, we have shown for the first time that supernematic
liquid-crystalline phases exhibit unusually large self-diffusivity which can
be explained in terms of a substantially reduced $\mathcal{S}_{\mathrm{rc}}$.
Here the suppression of $\mathcal{S}_{\mathrm{rc}}$ is assisted by confinement
to a nanoscopic slit-pore where the solid substrates favor a planar
arrangement of the molecules with respect to the substrate plane. Because the
solid surfaces of the slit-pore may be viewed as the representation of an
external field superimposed onto the intermolecular interactions, we
anticipate the results presented here to be generic in that they should
persist in other fluids composed of anisometric molecules that are exposed to
external fields (e.g., dipolar fluids in external magnetic fields). Hence, we
believe our results to be important for a broad range of liquid-crystal
applications ranging from lubricants in
nanotribology \cite{carrion09} over nanosensors \cite{hussain09} to photonic \cite{juodkazis09} and organic
electronic devices \cite{yang09} where the mobility of molecules plays a key
role.

Though the experimental data so far communicated in the literature do not yet provide a rigorous confirmation of our theoretical predictions, some of these data are in reasonable qualitative agreement with our simulations. For example, extrapolating longitudinal relaxation rates reported in Ref.~\citealp{miyajima84} from the RN to the N phase yields values markedly below those in the N phase. If referred to the same $T$ these relaxation rates correspond to correlation times that are notably shorter in the RN compared with the N phase. However, this general interpretation of NMR data remains speculative as long as translational diffusion has not definitely be identified as the process governing the observed relaxation. 

Direct evidence for the relation between translational diffusion and NMR data can be provided by the pulsed field gradient NMR (PFG NMR) technique \cite{valiullin09} which records molecular displacements typically over a $\mu$m range. For example, PFG NMR has been applied to directly assess the diffusion tensor upon entering the N phase \cite{dvinskikh01}. In these studies diffusion in the direction of the molecules' long axes was found to increase with increasing nematic order. By the same technique the diffusivity of n-alkanes in nanochannels was found to increase with increasing orientational order \cite{valiullin06}. These findings are in line with our data where enhanced molecular ordering is accompanied by increasing diffusivities in the direction of $\widehat{\bm{n}}$. The powerful combination of PFG NMR with magic angle spinning has recently enabled a notable increase in both observation times and gradient pulse intensities \cite{maas96}. As a consequence PFG NMR diffusion measurements became possible beyond the limits of measurability existing so far. This concerns in particular the first diffusion measurements with liquid crystals confined to nanopores \cite{romanova09}. To stimulate a direct experimental verification of our present predictions using these novel techniques is the primary purpose of this study.

\acknowledgments Financial support from the International Graduate Research
Training Group 1524 is gratefully acknowledged.

\end{document}